\documentstyle[aps,amssymb,epic,preprint]{revtex}

\def\tr{{\rm Tr\, }}
\def\c{{\Bbb C}}

\def\dd{\hbox{\kern0.3em/\kern-0.7em /\kern0.5em}}
\def\lie#1{{\rm Lie}\left( #1 \right)}

\def\df{$D-$flat }

\def\a{\alpha}
\def\b{\beta}

\def\hp{\hat{\phi}}

\def\u{\overline{u}}
\def\v{\overline{v}}

\def\q{\tilde{Q}}

\def\ssqr#1#2{{\vbox{\hrule height #2pt
\hbox{\vrule width #2pt height#1pt \kern#1pt\vrule width #2pt}
\hrule height #2pt}\kern- #2pt}}

\newcommand{\drawsquare}[2]{\hbox{%
\rule{#2pt}{#1pt}\hskip-#2pt
\rule{#1pt}{#2pt}\hskip-#1pt
\rule[#1pt]{#1pt}{#2pt}}\rule[#1pt]{#2pt}{#2pt}\hskip-#2pt
\rule{#2pt}{#1pt}}

\newcommand{\Yfund}{\raisebox{-.5pt}{\drawsquare{6.5}{0.4}}}
\newcommand{\Ysymm}{\raisebox{-.5pt}{\drawsquare{6.5}{0.4}}\hskip-0.4pt%
        \raisebox{-.5pt}{\drawsquare{6.5}{0.4}}}
\newcommand{\Ythrees}{\raisebox{-.5pt}{\drawsquare{6.5}{0.4}}\hskip-0.4pt%
          \raisebox{-.5pt}{\drawsquare{6.5}{0.4}}\hskip-0.4pt%
          \raisebox{-.5pt}{\drawsquare{6.5}{0.4}}}
\newcommand{\Yasymm}{\raisebox{-3.5pt}{\drawsquare{6.5}{0.4}}\hskip-6.9pt%
        \raisebox{3pt}{\drawsquare{6.5}{0.4}}}
\newcommand{\Ythreea}{\raisebox{-3.5pt}{\drawsquare{6.5}{0.4}}\hskip-6.9pt%
        \raisebox{3pt}{\drawsquare{6.5}{0.4}}\hskip-6.9pt
        \raisebox{9.5pt}{\drawsquare{6.5}{0.4}}}

\begin{document}
\tightenlines


\preprint{\vbox{
\hbox{hep-th/9905052}
}}
\title{Non Supersymmetric Vacua and the D-flatness Condition}
\author{Gustavo Dotti \thanks{Fundaci\'on Antorchas postdoctoral fellow.}}
\address{FaMAF, Universidad Nacional de C\'ordoba,\\
Ciudad Universitaria, 5000, C\'ordoba, ARGENTINA}
\date{October 1998}

\maketitle

\begin{abstract}
Some  $N=1$ gauge theories, including SQED and $N_F=1$ SQCD,  
have the property that, for arbitrary superpotentials, 
all stationary points of 
the potential $V=F+D$ are $D-$flat. For others, stationary points 
of $V$ are  complex gauge transformations of \df configurations.
As an implication, the  technique to parametrize the moduli 
space of supersymmetric vacua in terms of a set of basic 
holomorphic $G$ invariants can be extended to non-supersymmetric vacua.  A similar situation is found 
in non-gauge theories with a compact global symmetry group.
\end{abstract}


\section{Introduction} \label{intro}

One interesting feature of supersymmetric gauge theories 
is the existence of multiple, physically inequivalent,  
$V = 0$ vacua~\cite{is}. This brings the notion of ``moduli space" 
${\cal M}_{sv}$ 
of supersymmetric vacua (sv), the set of sv of a  
theory mod $G$ 
transformations, $G$ the gauge group of the theory. 
Classically, there is a well known construction of 
${\cal M}_{sv}$~\cite{ps,lt}. Let $\c ^n = \{ \phi \}$ 
be the  vector space of constant matter field configurations, 
$\hp^i(\phi), i=1,...,s$ a basic set of holomorphic $G$ invariants, 
${\cal D} \subseteq \c^s$ the algebraic subset of $\c^s$ defined 
by the polynomial 
constraints among the basic invariants. 
There 
is precisely one closed orbit of the complexification $G^c$ of 
the gauge group in each level set $\hp ^i (\phi) = \hp ^i _0$,  and  
 there is a unique $G$ orbit of \df points per closed $G^c$ orbit
(no \df point can be found in  non-closed $G^c$ orbits). 
Thus ${\cal D}$ is the moduli 
space of \df points, and ${\cal M}_{sv}$ is the subset of 
${\cal D}$ selected by the condition $\partial W = 0$. 
For some theories the above picture changes drastically in the 
quantum regime, where all sv are lifted~\cite{is}.  
For others, the quantum moduli space of sv is the same as 
${\cal M}_{sv}$~\cite{css}, or a deformation of ${\cal M}_{sv}$ 
in its ambient vector space $\c^s$~\cite{is,gn}. 
In the latter case, knowledge of ${\cal M}_{sv}$ plays a crucial 
role in the determination of the quantum moduli space of sv.\\
 
In  this work  we study {\em non supersymmetric} vacua 
(nsv) in the classical regime, as a first stage in the understanding 
of nsv in the quantum regime. A first look at the problem suggests 
that  no much can be said about nsv, here defined to be 
$V \neq 0$ local minima of the scalar potential $V$. 
Firstly, there are 
strong limitations on a gauge or non-gauge supersymmetric 
theory to admit nsv. As an example, dimensionful constants 
are required in the superpotential $W(\phi)$ to allow 
terms with different powers of fields, otherwise  
$W(\phi)$ would be a homogeneous function 
on the chiral fields $\phi$, $W(x\phi)=x^d W(\phi)$, 
and every  stationary point $\partial V = 0$ 
would be  a sv, as  $0=\phi 
\partial V / \partial \phi = (d-1)F \; (+2D)$. 
Secondly, for theories with  nsv,  
there does not seem to be any reasonable way 
to parametrize its moduli space ${\cal M}_{nsv}$. 
Once the $D-$flatness condition 
is removed we may expect  nsv in non-closed $G^c$ orbits. 
The basic holomorphic invariants do not  separate $G^c$ orbits, 
they are only able to ``distinguish" two different 
$G^c$ orbits if they are closed. 
We could tackle 
this problem by using the techniques developed in~\cite{as} 
to find the extrema of functions which are invariant under 
the action of a compact Lie group $G$. The $G$ orbits 
are the level sets  
of a {\em complete}  (holomorphic and non-holomorphic) basic set of 
$G$ invariants $\psi^j(\phi, \phi^{\dagger}), j=1,...,k$. The 
$\psi^j$'s are subject to polynomial (in)equalities that 
define a semi-algebraic subset ${\cal O}$ of  ${\Bbb R}^{2s} \simeq 
\c^s$~\cite{as}. 
The  extrema 
of $G$ invariant functions can be found by working directly in  
the orbit 
space  ${\cal O}$~\cite{as}.  
However, computations are cumbersome because  a detailed 
knowledge 
of the $G$ strata in ${\cal O}$ is required.
 In this work we explore a simpler alternative which is 
based on the simple structure of the scalar potential $V = F + D$.
Note that 
 $F$  is the square norm of the  $G^c$ ``covector" 
(i.e., transforming as $\overline{\rho}$ if 
$\phi$ is in the $\rho$ representation) $\partial W$, 
whereas 
 $D$  is the square norm of the field $\Phi^A = \phi^{\dagger} T_B \phi 
K^{BA} \in \lie G$, $K^{BA}$ the inverse Killing metric 
in $\lie G$.~\footnote{$\Phi^A$ transforms as an adjoint field under $G$ but  
this picture breaks after complexifying $G$.} 
For a large set of groups and representations this structure 
of $V$ restricts {\em all}  
stationary points of $V$  (not only sv) to closed $G^c$ orbits.
This fact  not only simplifies the search of nsv, 
it also allows to  construct the moduli space 
${\cal M}_v$ of {\em all} vacua, supersymmetric 
and non-supersymmetric, as a subset of ${\cal D}$, 
i.e., ${\cal M}_{sv} \subseteq {\cal M}_v \subseteq {\cal D}$.\\

In non gauge theories with a global symmetry group $G$ the 
scalar potential equals the square norm  $|\partial W|^2$
of the $G^c$ covector $\partial W$. For a large set of groups 
and representations  
this implies that  
nsv are restricted to closed $G^c$ orbits, i.e., they are $G^c$ related 
to (formal)  \df points $\phi^{\dagger}T\phi = 0$ for all $ 
T \in \lie G$. Thus, the $D-$flatness condition plays a r\^ole 
in the search of nsv of theories with a {\em global} symmetry $G$!
Such ${\cal N} = 1$ theories  arise as the  low 
energy effective actions of confining gauge theories, and they often 
break supersymmetry. A well known example is the chiral theory 
with one flavor of matter in the four dimensional representation 
of  $SU(2)$~\cite{iss}. \\
The fact that   nsv occur only in closed $G^c$ orbits 
guarantees the exact ``doubling" of Goldstone bosons~\cite{s}.
We have doubling when  
${G^c}_{\phi}$, the little group of $G^c$ at the vacuum $\phi$, 
is the same as ${G_{\phi}}^c$, the complexification of the little 
group of $G$ at $\phi$ (in general, ${G_{\phi}}^c \subseteq 
{G^c}_{\phi}$, see~\cite{gs}). An equivalent condition is that 
 $T^{\dagger}$ be  unbroken 
whenever  $T \in \lie {G^c}$ 
is unbroken.\footnote{To see the equivalence write  
 $T  = (T+T^{\dagger})/2 + i(T-T^{\dagger})/(2i)$.} 
This condition is satisfied if  
the orbit $G^c \phi$ is closed, i.e.,   
if $\phi$ is $G^c$ related to a \df point. To 
show  this we can assume  that  $\phi$ is $D-$flat, 
as the  
$G^c$ isotropy groups of two points in a $G^c$ orbit are $G^c$ conjugated.
If $\phi$ is \df   
\begin{equation}
|T^{\dagger} \phi|^2 = \phi^{\dagger} T^{\dagger} T \phi + 
\phi^{\dagger} [T,T^{\dagger}] \phi, 
\end{equation}
then  
  $T^{\dagger} \; \in \lie {{G^c}_\phi}$ if  $T \in \lie {{G^c}_{\phi}}$. 
We should remark that the condition that $G^c \phi$
be closed for nsv   $\phi$ 
is {\em stronger}  than ${G_{\phi}}^c = {G^c}_{\phi}$. \\
The organization of this paper is as follows: in Section~\ref{sv} 
we introduce the notion of {\em fibers}, review the construction 
of ${\cal M}_{sv}$, and state the Hilbert-Mumford criterion 
for non-closed $G^c$ orbits; in Section~\ref{global} we 
study nsv of theories with a global symmetry. Section~\ref{gauge} is 
devoted to gauge theories, and includes a subsection on 
abelian gauge groups, for which a more systematic treatment is possible.
The main results are Theorem~I in Section~\ref{global} and 
Theorems~II and III in Section~\ref{gauge}. 

\section{Preliminaries} \label{sv}
Let $G$ be a compact, connected group, $\rho$ a unitary 
representation of $G$ on $\c^n$.  
  We will consider simultaneously the cases where 
 $\c^n = \{(\phi^1,\cdots,\phi^n) \}$  is the constant chiral field
configuration space 
of a supersymmetric theory with global symmetry $G$, 
 or  the  
constant matter  chiral field configuration space  of an ${\cal N} = 1$ gauge
 theory, $G$ being the gauge group. Any $G$ invariant holomorphic 
polynomial $p(\phi)$ can be written in terms of a basic set 
of invariants ${\hp}^i(\phi), i=1,\cdots,s$ as 
\begin{equation}
p(\phi) = \hat{p}(\hp ^1(\phi),...,\hp ^s (\phi)), 
\end{equation}
where $\hat{p}$ is a polynomial $\c^s \to \c$ function~\cite{pv}.
In general, the basic invariants are constrained by polynomial equations 
$C^{\a}(\hp)=0$, meaning that $C^{\a}(\hp(\phi))\equiv 0$. 
The zero set ${\cal D} = \{\hp \in \c^s : C^{\a}(\hp)=0 \} \subseteq 
\c^s$ plays an important r\^ole in the construction of the 
moduli space of supersymmetric vacua of the gauge theory with 
matter content $\phi$ and gauge group $G$. 
This construction is better understood if we introduce the  
notion of ``fibers". Fibers are the 
level sets ${\hp}^i(\phi)={\hp}^i_0, i=1,\cdots,s$ of the basic 
invariants, they are closed, disjoint sets.  
The configuration space $\c^n = \{\phi \}$ 
is partitioned  into fibers, 
and the set of fibers
is parametrized by ${\cal D}$. 
 Every fiber contains 
complete orbits of the complexification $G^c$ of $G$, 
possibly infinitely many of them, only one 
of which is closed (in the topological sense)~\cite{ps}. 
The only 
closed $G^c$ orbit in a fiber $f$ lies in 
the boundary of any other $G^c$ orbit in $f$, and can therefore be found 
by taking the intersection of the closures of the $G^c$ orbits in $f$.  
Let  $T_A$ be  
a basis of hermitian generators of $G$ in the $\rho$ 
representation. A $G$ element admits the expansion $g=\exp(iC^AT_A)$ 
with real $C^A$'s, whereas a $G^c$ element admits a similar expansion with 
complex $C^A$'s. It follows that the $G^c$ action on $\c^n$ is non unitary.  Consider 
a ``pure imaginary" $G^c$ one dimensional  subgroup $g(s)=\exp(sT)$, 
$T$ a hermitian $\lie G$ generator (note the absence of the $i$ factor 
in the exponent) acting on an arbitrary $\phi \in \c^n$, and define  
 $\phi(s)\equiv g(s)\phi$, then~\cite{lt,kn} 
\begin{eqnarray} \label{d1}
\frac{d}{d s} \left( \phi^{\dagger}(s) \phi(s) \right) &=& 
2 \phi^{\dagger}(s) T \phi(s), \\ 
\frac{d^2}{d s^2} \left( \phi^{\dagger}(s) \phi(s) \right) &=&
4 (T \phi(s))^{\dagger} (T \phi(s)) \geq 0, \label{d2}
\end{eqnarray}
equality holding only  when 
$T$ is a generator of the 
little group $G_{\phi}$ of $\phi$ (and so $\phi(s) = $ constant). 
If $T \notin \lie {G_{\phi}}$, 
$\phi^{\dagger}(s) \phi(s)$ is a convex (positive second 
derivative) function of $s$. Convex ${\Bbb R} \to {\Bbb R}$ functions 
$f(s)$ are easily 
seen to satisfy the following three properties: (i) there is 
at most one stationary point of $f$; (ii) if $s_0$ 
is a stationary point of $f$, then it is a global minimum; 
(iii) if $f' \geq  0$ at  some point, 
then $\lim_{s \to \infty} f(s) = +\infty$.
From these properties, eqns.(\ref{d1},\ref{d2}) 
and Cartan's decomposition  $G^c = G{\cal T}G$, ${\cal T}$ a 
pure imaginary maximal torus, 
 follows that \df points  $\phi_D^{\dagger}T\phi_D=0$
are vectors of minimum length in a $G^c$ orbit, and that there is 
at most one $G$ orbit of such vectors in a given $G^c$ orbit.
It was found in~\cite{kn} that 
closed  $G^c$ orbits contain 
a unique $G$ orbit of \df points~\cite{kn}, that we will refer to as  the 
``core" of the $G^c$ orbit, whereas  no \df point 
can be found in a non closed $G^c$ orbit. These facts 
 allow a gauge independent 
characterization of 
 the $D-$flatness condition found in Wess-Zumino gauge: 
the supersymmetric vacua  of a gauge theory with gauge group $G$ 
lie on closed $G^c$ orbits. They  also allow to regard 
the set of fibers ${\cal D}$ as the  set of closed $G^c$ orbits, 
or the moduli space of \df points, i.e., the set of \df configurations 
mod $G$ transformations.  
The relevance to supersymmetric gauge theories 
of the connection between \df  configurations, minimal length 
vectors  and closed $G^c$ orbits found in~\cite{kn}
was  first pointed out in~\cite{ps}. 
The supersymmetric vacua (sv) of an ${\cal N} = 1$ gauge theory 
satisfy two conditions: (F) the $F-$flatness condition $\partial W = 0$ 
and (D) the $D-$flatness condition $\phi^{\dagger}T\phi=0 \, \forall 
T \in \lie G$. Condition (F) is $G^c$ invariant, 
every point in the orbit $G^c \phi_F$   of  an $F-$flat point $\phi_F$  
is $F-$flat, and, by continuity, every point in the closure 
$\overline{G^c{\phi}}$ is $F-$flat.
Condition (D) imposes an additional restriction: the sv lie on 
the core of   closed $G^c$ orbits. However, once an $F-$flat point 
$\phi_F$ is found, we know there is a $G$ orbit of sv 
in $\overline{G^c \phi_F}$, namely, the core of \df points in the only 
closed $G^c$ orbit in  $\overline{G^c \phi_F}$.
In other words, (F) selects the fibers $f$ where the sv live, (D) their  
location in $f$.
As there is  one closed $G^c$ orbit per fiber, which contains 
precisely one $G$ orbit of \df points, the moduli space 
of sv ${\cal M}_{sv}$  (sv mod $G$ transformations), 
is the same as the set of  fibers containing $\partial W = 0$ 
$G^c$ orbits. 
 ${\cal M}_{sv} \subseteq {\cal D}$ can be parametrized   
by 
adding to the constraint equations $C^{\a}(\hp)=0$ defining ${\cal D}$ 
the   $G$ invariant holomorphic equations resulting from 
$\partial W = 0$~\cite{lt}. In the special case $W = 0$,  
${\cal M}_{sv} = {\cal D}$, the moduli space of \df points. \\
  In  non-gauge theories with a global symmetry $G$, the sv satisfy 
only the $G^c$ invariant $F-$flatness condition.  Generically,  there 
are infinitely many $G$ orbits per $G^c$ orbit, and so there is no 
clear way to parametrize the moduli space of sv in non-gauge theories.   \\
In the following sections we show that, for a large set of gauge theories,  
the $V \neq 0$ 
stationary points of 
the scalar potential $V = F + D$, $F=|\partial W|^2, D = \frac{g^2}{8} 
\sum_A (\phi^{\dagger} T_A \phi)^2$, lie all on closed $G^c$ orbits  
(not necessarily in their cores), there being at most one $G$ orbit 
of stationary points of $V$ in a closed $G^c$ orbit. This leads to 
a parametrization of the moduli space ${\cal M}_{nsv}$ of nsv 
as a subset of ${\cal D}$, the set of closed $G^c$ orbits. 
${\cal M}_{nsv}$ is obtained by projecting onto ${\cal D}$ the 
stationary point condition $\partial V = 0$
and the condition that the boson mass matrix $\partial_i \partial _j V$ 
at the stationary point be positive semidefinite. This may result 
in non-holomorphic (in)equalities. 
The moduli space of vacua is then ${\cal M}_v = {\cal M}_{sv} 
\cup {\cal M}_{nsv} \subseteq {\cal D} \subseteq \c^s$. 
A similar situation is found 
in some  non-gauge theories with a global symmetry $G$,
their nsv are restricted to closed orbits of the complexification 
$G^c$ of the global symmetry group $G$, i.e., they  
are $G^c$ related to formal \df points. 
We make use of a theorem due to Mumford that says that, given a non-closed 
orbit $G^c \phi_0$, the closed $G^c$ 
orbit lying in the boundary of $G^c \phi_0$   
can be reached by means of a one dimensional pure imaginary  subgroup 
of $G^c$:\\
Theorem [Mumford]~\cite{gs,pv}: 
Assume $G^c \phi_0$ is not closed, then there is a 
hermitian generator $T$ of $G$  
 such that $\lim_{s \to \infty} 
\exp(-s T) \phi_0 = \phi_c$, and $G^c \phi_c$ is closed.  \\
Remark: if $\phi_0 = \sum_{\mu} {\phi_0}_{\mu}$ 
is the weight decomposition of $\phi_0$ (${\phi_0}_{\mu} \neq 0$), 
then $\mu(T) \geq 0 \; \forall \mu$ 
(and strictly positive for some $\mu$). This implies  
$|\phi_c| < |\phi|$, and also   $\lim_{s \to \infty} 
|\exp(s T) \phi| = \infty$.\\

\noindent
{\it Example \ref{sv}.1:}
 Consider $G=U(1)$ acting  on $\c^2$, $\phi = (u,v)$, 
$u$ a charge $1$ field and $v$ a charge $-1$ field. 
$\lie G = \text{span} (T), T = \text{diag}(1,-1)$.  
$G^c = GL(1,\c)$  
acting by $x \cdot (u,v) = (xu, x^{-1}v)$.
The set of basic invariants contains a single field $z = 
uv$, then ${\cal D} = \c^1$. The fibers $uv
=  z_0 \neq 0$ contain a single (therefore closed) 
$G^c$ orbit, with a core of vectors of 
minimum length (\df 
  points) satisfying $ uv = z_0, 
\; |u| = |v|$. The fiber $z = 0$ contains 
the closed orbit ${\cal O}_1 = \{(0,0)\}$ and the non-closed orbits 
${\cal O}_2 = \{(u,0), u \neq 0 \}, 
{\cal O}_3 = \{(0,v), v \neq 0 \}$, which do not 
contain vectors of minimum length. Also ${\cal O}_1 \subseteq 
\overline{{\cal O}_1} \cap \overline{{\cal O}_2} \cap \overline{{\cal O}_3}$. 
For points  in  ${\cal O}_2$ (${\cal O}_3$), 
$e^{-sT}$  ($e^{-s(-T)}$) is 
as in Mumford's theorem. If the $U(1)$ symmetry is local, 
and we add a superpotential $W(z)$ to this gauge theory, 
the sv condition  $0= \partial W = W'(z)(v,u)$ yields a single 
 holomorphic $G$ 
invariant equation, namely $zW'(z)=0$. This equations 
selects the fibers containing $\partial W = 0$ $G^c$ orbits. 
As there is a unique $G$ orbit of \df points per fiber,
 the moduli space 
of sv of this gauge theory  is ${\cal M}_{sv}=\{z \in \c | zW'(z)=0\}$. 
If the $U(1)$ symmetry were global, every point in  fibers 
$z_0$ satisfying $W'(z_0) = 0$ 
would be a sv. As every fiber 
contains infinitely  many $G$ orbits, there is no clear 
way to parametrize ${\cal M}_{nsv}$.    \\

\noindent
{\it Example \ref{sv}.2:} Consider a theory with a matrix $M$ of chiral 
fields and a superpotential invariant under $M \to g M g^{-1}, 
g \in SU(N)$. The configuration space is $\c^{N^2}$, $G=SU(N)$, 
$\rho = adj\; + \; {\bf 1}$, and $G^c=SL(N,\c)$. The adjoint field is 
 $A^{\a}_{\b} = M^{\a}_{\b} - \frac{1}{N} \delta^{\a}_{\b} \tr M$, and 
the singlet is $u = \tr M$.  
The holomorphic 
invariants are ${\hp}^1 = u$ and ${\hp}^i = \tr A^i, i=2,\cdots,N$, 
they are unconstrained and so ${\cal D} = \c^N$.  
Jordan's decomposition 
implies that in every $G^c$ orbit  there is an element of the form $(u,A)$, 
$A = S + N$, where $S$ is diagonal, $N$ strictly upper triangular, 
and $[S,N]=0$,  
these are the semisimple and nilpotent parts of $A$. Note that 
 $\hp ^i  = \tr S^i, i>1$ then $(u,S+N)$ and $(u,S+N')$ belong 
to the same fiber. 
In~\cite{pv}, section 8.5,  it is established that the $G^c$ orbit of 
$S+N$ is closed iff $N=0$. 
As there is one closed $G^c$ orbit per fiber we conclude 
that if $S$ and $S'$ are semisimple 
and  $\tr S^i = \tr S'^i, i=2,...,N$ then $S'=gSg^{-1}, g \in SL(N,\c)$. 
As there is a finite number of $G^c$ orbits of nilpotent $A's$~(~\cite{pv}, 
section 8.5)  
every fiber $(u,\tr A^i)=(u_0,\hp ^i_0)$ 
contains the same (finite) number of $G^c$ 
orbits, a picture that differs substantially from that 
of Example~II.1.  Mumford's curve 
``switches off" the nilpotent piece of the adjoint field.  
Take, e.g., $N=3$, $\phi_0=(A_0,u)$, 
$A_0 = S + N$,  
\begin{equation}
A_0 = \left( \begin{array}{rrr}
 -2 & 0 & 0 \\ 0 & 1 & 1 \\ 0 & 0 & 1 \end{array} \right), \;\; 
S = \left( \begin{array}{rrr}
 -2 & 0 & 0 \\ 0 & 1 & 0 \\ 0 & 0 & 1 \end{array} \right), \;\; 
N = \left( \begin{array}{rrr}
 0 & 0 & 0 \\ 0 & 0 & 1 \\ 0 & 0 & 0 \end{array} \right), \;\; 
\end{equation}
A choice of $T$ satisfying Mumford's theorem is 
\begin{equation} 
T = \left( \begin{array}{rrr}
 0 & 0 & 0 \\ 0 & 1 & 0 \\ 0 & 0 & -1 \end{array} \right). 
\end{equation}
Note that $\lim_{s \to \infty} \exp(-sT) A_0 = S$ and 
the square length  $\tr {(A_0^{\dagger}A_0)} = 7 > \tr {(S^{\dagger}S)} = 6.$
Consider the gauge theory with superpotential 
 $W = uA^{\a}_{\b}A^{\b}_{\a} 
+ m u^2/2 + \gamma u \equiv  u \hp  + m u ^2/2 + \gamma u \; 
(\hp \equiv \hp ^2)$, 
$\partial W = (2uA,\hp +mu+\gamma) =0$ iff 
(i) $u=0$ and $\hp=-\gamma$ or (ii) $A=0$ and $u=-\gamma/m$. 
The  fibers containing sv (in the cores of their closed 
$G^c$ orbits) are (i) $\hp ^1=0, \hp ^2 = -\gamma$ and 
arbitrary $ \hp ^j, j \geq 3$, and (ii) $\hp ^1 = - \gamma/m, 
\hp ^j = 0, j \geq 2$, thus ${\cal M}_{sv} = \{ (\hp ^1,\cdots, 
\hp ^s) \in \c ^s | \hp ^1=0, \hp ^2 = -\gamma \} \cup 
 \{ (\hp ^1,\cdots, \hp ^s) \in \c ^s | \hp ^1 = - \gamma/m, 
\hp ^j = 0, j \geq 2 \}$. Again, if the $SU(N)$ symmetry were global, 
${\cal M}_{sv}$ constructed above would {\em not} be a parametrization of  
the moduli space of sv, as there are infinitely many $G$ orbits 
in each $\partial W=0$ $G^c$ orbit of type (i).

\section{Non-supersymmetric vacua in theories with a global symmetry} \label{global}
If $W$ is a $G$ invariant superpotential 
 its gradient $\partial W$ transforms as a $G^c$ ``covector"
\begin{equation}
\label{cov}
W(g\phi)=W(\phi), \;\; \partial W(g\phi)=\partial W(\phi) g^{-1}.
\end{equation} 
It is useful to think of $\partial W (\cdot)$ as a map $\c^n \to {\c^n}^*$ 
commuting with the $G$ actions $\rho$ and $\overline{\rho}$.
The vector $\phi$ is assigned the covector $\partial W (\phi)$,  
 $F = |\partial W|^2$ measures its square length. 
It follows from~(\ref{cov}) that under this map 
 the orbit  $G^c \phi = \{g \phi | g \in G^c \} \subseteq  \c^n$
 gets mapped onto the orbit $G^c \partial W (\phi) \subseteq  {\c^n}^*$; 
also $G_{\phi} \subseteq G_{\partial W(\phi)}$, 
 $G_{\partial W(\phi)}$ being the little group of the 
${\c^n}^*$ point $\partial W (\phi)$, $G_{\phi}$ the little group 
of $\phi$~\cite{gs}.
We  exploit the fact that eqs.~(\ref{d1}, \ref{d2}) and all the 
results of the previous section apply to 
{\em any} $G$ representation, in particular   $\overline{\rho}$,  
 where $\partial W$ lives. Thus, if $F(\phi_0)$ 
is a local minimum of $F$, $\partial W(\phi_0)$ is a covector 
of minimum length in its $G^c$ orbit, then $G^c \partial W(\phi)$ 
must be  closed, 
and $\partial W(\phi_0)$ satisfies the $*D-$flatness condition  
\begin{equation} \label{*} 
(\partial W (\phi_0)) (-T) (\partial W (\phi_0))^{\dagger} = 0, 
\; \forall T \in \lie G. 
\end{equation} 
We prove now that, 
under certain assumptions, this implies that 
$G^c \phi_0$ itself is closed. 
To see this, define for any $\phi_0$ and hermitian $T$ the curve
$\phi(s) \equiv e^{-sT}\phi_0$ and also 
$F(s) \equiv [\partial W(\phi(s))] [\partial W(\phi(s))]^{\dagger} 
= |(\partial W(\phi_0)) \exp(s T)|^2$. Applying 
 (\ref{d1},\ref{d2}) to the $\overline{\rho}$ representation (or 
just computing the second derivative  of $F(s)$) we  see 
that,  whenever $T \notin \lie {G_{\partial W(\phi_0)}}$,    
$F(s)$ is a convex ${\Bbb R} \to {\Bbb R}$ function. 
If $\partial F(\phi_0)=0$, 
then  
$0=F'(0)=\partial W(\phi_0) (-T) (\partial W(\phi_0)) ^{\dagger}$,  
$F(0)$ is a global minimum of $F(s)$, 
and  $\lim_{s \to \pm \infty}F = \infty$.
As a consequence $G^c \phi_0$ must be closed.  If it were not, 
we could choose $T$ as in Mumford's 
theorem and get to a contradiction: 
$F(\phi_c) = \lim_{s \to \infty} F(s) = \infty,$ where 
 $\phi_c = \lim_{s \to \infty}
 \phi(s)$~\footnote{Even if $W$ 
has singularities, it is not possible 
that $F$ be well defined at $\phi_0$ and singular at $\phi_c$. 
This is so because one can always write $W(\phi) = W 
(\hat \phi(\phi))$, 
then $\partial W = (\partial  W / \partial \hat \phi ^j) 
\partial \hat \phi ^j$. 
 Now $\partial W / \partial \hat \phi ^j$ is constant on $G^c \phi_0$ 
and the $\partial \hat \phi ^j$ are polynomials, so no singularity can 
develop along the bounded $\phi(s), s \geq 0$ curve.} 
We conclude that $G^c \phi_0$ being non-closed forbids   $\phi_0$ 
from being   a stationary 
point of $F$. The only exception is when, 
for any $T$ as in Mumford's theorem, 
$T \in \lie {G_{\partial W(\phi_0)}}$. 
If this is  the case then $F$ is non-confining, 
that is  $\lim_{s \to \infty}|\exp (sT)\phi_0| = \infty $ while  
 $\lim_{s \to \pm \infty} F(\exp (sT)\phi_0)=F(\phi_0) < \infty$.
For $\phi_0$ and $T$ as in Mumford's theorem
 the weight decomposition $\partial W(\phi_0) = \sum_{\lambda} 
(\partial W(\phi_0))_{\lambda}$ is such that $\lambda (T) \leq 0 \; 
\forall \lambda$, then $F(\phi_c) < F(\phi_0)$ except 
in the non-confining case $\lambda (T) = 0 \; \forall \lambda$, 
where $F(\phi_c)=F(\phi_0)$.\footnote{If this is the case, 
and we are only interested in the spectrum of vacuum energies, we 
can use the fact that $F(\phi_0) = F(\phi_c)$ 
and still  restrict the search of vacua to closed $G^c$ orbits.}  
These observations are gathered in the 
following theorem:\\

\noindent
{\bf Theorem I:} Assume $G^c \phi_0$ is non-closed and $\phi_c$ is 
as in Mumford's theorem. \\
(a) $F(\phi_c) \leq F(\phi_0)$, a lower energy 
point can be found in the closed $G^c$ orbit in the boundary 
of $G^c \phi_0$.\\ 
(b) If  $G_{\phi_0} = G_{\partial W (\phi_0)}$ then: 
(i) $\phi_0$ cannot be a stationary point of $F$, 
(ii) $F(\phi_c) < F(\phi_0)$. \\
(c) Define 
$\widehat{\c^n}= \{ \phi \in \c^n | G_{\phi} = G_{\partial W(\phi)} \}$.
The moduli space ${\cal M}_{nsv}$ of non-supersymmetric vacua in  
$\widehat{\c^n}$ is the subset of ${\cal D}$ obtained by projecting 
onto ${\cal D}$ the  (in)equalities resulting from 
$\partial F =0$ and $\partial_i \partial_j F$ positive semidefinite.\\ 

To prove (c), note from  
(b) and the above discussion 
 that, in the sector $\widehat{\c^n}=\{\phi \in \c^n | 
G_{\phi} = G_{\partial W(\phi)} \}$ of the configuration space 
$\c^n$,  the stationary points $\phi_s$ of $F$ lie all on  closed 
$G^c$ orbits, satisfy the $*D-$flatness condition eq.(\ref{*}) 
and are global minima of the restriction of $F$ to 
$G^c\phi_s$ (in particular,  no local maximum 
of $F$ exists in $\widehat{\c^n}$).
 Moreover  there is at most one  
$G$ orbit of nsv per closed $G^c$ orbit. As the set of closed $G^c$ 
orbits is parametrized by ${\cal D}$, the moduli space of nsv 
in $\widehat{\c^n}$ is the subset of ${\cal D}$ obtained by projecting 
onto ${\cal D}$ the (in general non-holomorphic) (in)equalities 
resulting from the conditions 
$\partial F = 0$ and $\partial_i \partial_j F$ positive semidefinite.
Besides simplifying the search 
of nsv in $\widehat{\c^n}$, theorem~I shows a construction  
of ${\cal M}_{nsv}$  closely related  to the parametrization 
of ${\cal M}_{sv}$ in gauge theories. \\

\noindent
{\it Example \ref{global}.1:}
Consider the theory of example \ref{sv}.1 with the $U(1)$ 
symmetry global. $\partial W = W'(z)(v,u)$, 
then $U(1)_{\partial W(u,v)} = U(1)_{(u,v)}$ except 
at nonzero sv, i.e., $\widehat{\c^2}= \c^2 \setminus \{  (u,v) \neq (0,0) |
W'(uv) = 0 \}$. If such a vacuum exists, $F$ is non-confining, 
meaning that $F$ is constant along the $GL(1,\c)$ orbit 
of the nontrivial sv, which extends to infinity. 
Theorem~I guarantees that   the nsv lie all on closed $GL(1,\c)$ 
orbits, as they are all in $\widehat{\c^2}$. 
 In fact,    
 $F = |W'(z)|^2(u \u + v \v)$ and   $\partial F = 0$ yield 
$0 = u \partial F/\partial u 
- v \partial F / \partial v = |W'(uv)|^2(u\u  - v \v )$.
This means that every 
stationary point $(u,v)$ of $F$ in $\widehat{\c^2}    $ is  \df, 
and so its   $GL(1,\c)$ orbit is closed, as predicted. 
To construct the moduli space of nsv
we project $\partial F =0$ and $\partial ^2 F \geq 0$ 
onto ${\cal D}$. This is readily done if we replace 
$(u,v)$ in $\partial F = 0$ and $\partial ^2 F \geq 0$ 
by the  \df representative $u=v=\sqrt{z}$ in the $uv=z$ fiber.  
For details refer to example~\ref{gauge}.1, the result is
${\cal M}_{nsv} = \{ z \in \c^1 | W'(z)+2zW''(z) = W''+zW'''=0  \}$. \\

\noindent 
{\it Example \ref{global}.2:} Consider the  theory of example~\ref{sv}.2, 
with a {\em global} $SU(N)$ symmetry. 
$\partial W = (2uA,\hp+mu+\gamma) =0$ iff 
(i) $u=0$ and $\hp=-\gamma$ or (ii) $A=0$ and $u=-\gamma/m$. 
Condition (i) defines a fiber of sv containing non-closed $G^c$ orbits 
extending to infinity, i.e,  
$F$ is not confining and this  explains the existence of stationary $F$ points 
in non-closed orbits. 
In the $u \neq 0 $ sector $SU(N)_{(u,A)} = SU(N)_{\partial W(u,A)}$, 
therefore $\widehat{\c^{N^2}}     = \{(u,A) \in \c^{N^2} | u \neq 0 \} \cup 
\{(0,0)\}$. 
All $\partial W \neq 0$  stationary points of $F$  lie 
in the $u \neq 0, A \neq 0$ sector of the configuration 
space, where   
 Theorem I applies. In particular, these
stationary configurations must lie on closed $G^c$ orbits. In fact, from 
$0=\partial F / \partial A$ and $u\neq 0$ we obtain 
\begin{equation} A^{\dagger} = -A \frac{({\hp}^{\dagger}+
\overline{m} u^{\dagger} + \overline{\gamma}
)}{2uu^{\dagger}} \equiv -A e^{-i\a}, \end{equation}
from where $[A,A^{\dagger}]=0$, which implies $A$ is $SU(N)$ $D-$flat. Also 
$(\hp+mu+\gamma)/(2uu^{\dagger}) = e^{i \a}$, as this 
is an eigenvalue of the dagger 
operator. Adding $\partial F /\partial u=0$ we get the equations 
selecting the fibers containing  
$G$ orbits of stationary  points of $F$. There is only one such fiber:   
$u=xe^{i\a}/m, \; \hp=xe^{i\a}/2$; $e^{i\a}=\gamma/|\gamma|$ 
and $x=3m\overline{m} /8 -\sqrt{(3m\overline{m}/8)^2+|\gamma|
 m \overline{m}/2} < 0$.  \\

When proving (a) and (b) of Theorem~1 we showed that $G^c \phi$ 
is closed if $G^c \partial W (\phi)$ 
is closed and $\phi \in \widehat{\c^n}$ (the reciprocal 
requires $F$ to be confining in the sense described above). Yet, 
we should not expect  
the core of *\df points in $G^c \partial W(\phi)$ to be the image 
under $\partial W(\cdot)$ of the core of \df points in $G^c \phi$, 
a non-generic feature exhibited by the two previous examples. \\

\noindent
{\it Example \ref{global}.3:} Consider an $SO(N)$ theory with two vectors,  
$\vec{\phi_1}$ and $\vec{\phi_2}$, and a superpotential 
$W = \vec{\phi_1} \cdot (\vec{\phi_1}+i\vec{\phi_2})$. 
It can readily be checked that the isotropy groups $SO(N)_{\phi}$ 
and $SO(N)_{\partial W(\phi)}$ agree for every $\phi=(\vec{\phi_1},
\vec{\phi_2})$ in the configuration space $\c^{2N} = \widehat{\c^{2N}}$. 
If  $G^c \partial W(\phi)$ is closed, then so is $G^c \phi$.  
Moreover,  $G^c \phi$ is closed iff $G^c \partial W(\phi)$ is 
closed,
this superpotential also satisfies the confining condition. 
However, for  \df $\phi$, $\partial W (\phi)$ is not *\df in general. \\
 
\noindent
{\it Example \ref{global}.4:} 
Theorem 3.9 in \cite{ps} states that a point  $\phi_0$ 
is \df 
iff there is a holomorphic $G$ invariant $h(\phi)$ 
such that $\phi_0^{\dagger} 
= \partial h (\phi_0)$. In the special case where the set of basic invariants 
contains a single field $\hp(\phi)$ this theorem implies that any \df point 
satisfies the *$D$-flatness condition~(\ref{*}), 
as $\partial W = W'(\hp) \partial \hp$.
Write $\hp(\phi)=C_{(i_1 \cdots i_d)} \phi^{i_1} \cdots \phi^{i_d}$ and 
consider the $\c^n \to {\c^n}^*$ map  
$\phi^j \to \psi_i \equiv 
C_{(i i_2 \cdots i_d)} \phi^{i_2} \cdots \phi^{i_d}$.  
If $\rho$ is real then $d=2$,  ${C^{\dagger}}^{ik}C_{kj}=\delta^i_j$,  
$\partial_i W(\phi) = W'(\hp) C_{ij} \phi^j$, then 
$\widehat{\c^n} = \c^n \setminus \{\phi \neq 0 | W'(\hp(\phi))=0 \}$. 
Also 
$F = |W'|^2 \phi^{\dagger} \phi$, and $(\partial F) T \phi =  
|W'|^2 \phi^{\dagger}T \phi$. In the $\widehat{\c^n}$ sector 
stationary point are seen to lie {\em in the core} of closed $G^c$ 
orbits. This generalizes the situation of example III.1.

\section{Non-supersymmetric vacua in gauge theories} \label{gauge} 
 In many 
interesting  examples, the  $D$ term $\sum_A (\phi^{\dagger} T_A \phi)^2$ 
along the orbit of a pure imaginary one dimensional subgroup 
$\exp(-sT)$ of $G^c$ is a convex function of $s$, i.e., $d^2D(\exp(-sT)\phi_0)
/ ds^2  > 0 \, \forall s \in {\Bbb R}$. 
For $\phi_0$ and $T$  as in Mumford's theorem, this implies  
that $\phi_0$ cannot  be a stationary point of the scalar 
potential $V=F+D$, as $V'' \geq D'' > 0$. If it were, $V$ would diverge 
at $\phi_c = \lim_{s \to \infty}\phi(s)$. 
Assume there is a sector $\widehat{\c^n}$ of the configuration space 
where, for every point in 
non-closed $G^c$ orbits there is a choice of $T$ as in Mumford's theorem 
for which $d^2D/ds^2 > 0$ for all $s$. Stationary points of $V$ in $\widehat{\c^n}$ are  restricted 
to closed $G^c$ orbits. If also $d^2D(\exp(-sT)\phi_c)/ds^2 > 0$ 
for any $\phi_c \in \widehat{\c^n}$ 
in closed $G^c$ orbits and any $T \in (\lie G 
\setminus \lie {G_{\phi}})$, 
we can show, as in sections~\ref{sv} and~\ref{global}, 
 that there is at most one $G$ orbit of stationary 
points of $V$ per closed $G^c$ orbit.  The stationary point 
condition  $V'(0)=0$ reads
\begin{equation} \label{m} 
\partial W (-T) (\partial W)^{\dagger} 
+ \frac{g^2}{4}\phi^{\dagger}T_A \phi \phi^{\dagger}(TT_A+T_AT)\phi=0.
\end{equation} 
We gather the above observations in the following theorem:  
(in the aim of seeking simplicity
 we made some assumptions  stronger than necessary). \\

\noindent
{\bf Theorem II:} Restrict to the sector  $\widehat{\c^n}= 
\{ \phi \in \c^n \, | \, 
d^2D(\exp(-sT)\phi)/ds^2 >0 
\; \text{whenever} \; T \notin \lie {G_{\phi}} \}$ 
of the configuration space $\c^n$, then: \\
(a) For any superpotential, 
every stationary point $\phi_s$ of $V=F+D$ lies in a closed $G^c$ orbit, 
(equivalently, it is $G^c$ related to a \df configuration), 
satisfies 
the modified $D-$flatness ($MD-$flatness) condition eq.~(\ref{m}), 
and is a global 
minimum of the restriction of $V$ to $G^c \phi_s$. 
In particular, there is no 
local maximum of $V$.\\
(b) The moduli space of vacua ${\cal M}_{v}$ is the subset 
of ${\cal D} \subseteq \c^s$ obtained by adding to the constraint equations 
among basic invariants the non-holomorphic (in)equalities resulting from the 
stationary point condition $\partial V / \partial \phi = 0$ and the 
 condition that 
the boson mass matrix $\partial_i \partial_j V$ at the stationary point be positive semidefinite.\\

The proof of (b) follows again from the fact that there is 
at most one $G$ orbit of stationary points in a closed $G^c$ orbit 
and that ${\cal D}$ is the set of closed $G^c$ orbits. For supersymmetric 
vacua the projection of $\partial V = 0$ onto 
${\cal D}$ reduces to the $G$ holomorphic invariant equations 
obtained from $\partial W = 0$, and $\partial ^2 V \geq 0$ 
does not add any restrictions. ${\cal M}_v$  
is the union of the moduli spaces of sv and nsv, 
${\cal M}_v = {\cal M}_{sv} \cup {\cal M}_{nsv} \subseteq {\cal D}$.\\
 
\noindent
{\it Example \ref{gauge}.1:} Following the notation of examples
 \ref{sv}.1 and \ref{global}.1,  the $D$ 
term of SQED is $D = (u\u - v \v)^2 = |\phi|^4 - 4 |z|^2, \phi=(u,v)$.
As $z$ is $G^c$ invariant, $|z|^2$ is a constant along any 
$\phi(s) = \exp(-sT)\phi$ curve, whereas $|\phi(s)|^4$ 
is clearly a convex function (whenever $T \notin \lie {U(1)_{(u,v)}}$), 
and so is $D(s)$.
Alternatively, we can apply  eqs~(\ref{d1}, \ref{d2}) to the 
1-dimensional charge 
$2,-2$ and $0$ $U(1)$ representations $u^2$, $v^2$ and $uv$ to show that 
$D = |u^2|^2 + |v^2|^2 - 2 |uv|^2$ is the sum of two convex functions 
and a constant. In this example the configuration 
space $\widehat{\c^2}    $ equals $\c^2$, and Theorem~2 holds everywhere. 
Given an arbitrary $W(z)$, 
$V= |W'|^2(|u|^2+|v|^2) 
+ \frac{g^2}{8}(|u|^2-|v|^2)^2$. The stationary point condition 
$\partial V=0$ is always satisfied at the origin $\phi=0$ and 
at no other point in the $uv=0$ fiber. For 
nonzero $uv$ 
it is equivalent to $0 = u \partial V / \partial u \pm 
v \partial V / \partial v$:
\begin{eqnarray} \label{qed1} 
0 &=& \left(|W '|^2+\frac{g^2}{4}(|u|^2+|v|^2)\right) (|u|^2-|v|^2)  \\ 
0 &=& \overline{W '} (W '+2z W '')(|u|^2+|v|^2) + \frac{g^2}{4}
(|u|^2-|v|^2)^2. \label{qed2}
\end{eqnarray}
Eq.(\ref{qed1})  forces $D=0$, showing that stationary points 
lie on  closed $G^c$ orbits, as predicted.
Projecting~(\ref{qed2}) onto ${\cal D}$ we obtain the equations 
characterizing the fibers containing critical points, 
namely $0=zW'(z)(W'(z)+2zW''(z))$. 
To project  $\partial_i \partial_j V \geq 0$  
 at stationary 
points  onto ${\cal D}$ we use the section 
${\cal D} \ni z \to (u=\sqrt{z},v=\sqrt{z}) \in \c^2$. 
When replacing $u=v=\sqrt{z}$ and $W'(z)+2zW''(z)=0$ in 
the equations requiring that the eigenvalues of $\partial_i \partial_j V$ 
be $\geq 0$,   the inequalities reduce to $W''+zW'''=0$. Thus  
${\cal M}_v = \{ z \in \c^1 | zW'(z)=0 \} 
\cup  \{ z \in \c^1 | W'(z)+2zW''(z) = W''+zW'''=0  \}
= {\cal M}_{sv} \cup {\cal M}_{nsv}$. 
The equations defining ${\cal M}_{nsv}$ 
are independent of $g$, this is also the moduli space of nsv of 
the non-gauge theory of example \ref{global}.1. \\

As a first step towards generalizing  the ideas 
behind  the previous example  we 
re-write the $D$ term using  the  $G$ representation 
$\rho \otimes _s {\rho}$. Let $\phi = \sum_r \phi_r$ 
be the decomposition of $\rho$ into irreps, then
\begin{equation} \label{D0} 
D = \sum_{r,s} (\phi_r^{\dagger} T_A^r \phi_r) (\phi_s^{\dagger} 
T_A^s \phi_s) = \sum_{r,s} 
(\phi_r \otimes \phi_s)^{\dagger}(T_A^r \otimes T_A^s) 
(\phi_r \otimes \phi_s).
\end{equation}
 Using  $T_A^{r \otimes s} =  T_A^r \otimes {\Bbb I}_s 
+ {\Bbb I}_r \otimes T_A^s$ we obtain 
\begin{equation} \label{cas}
T_A^r \otimes T_A^s = \frac{1}{2}[(T_A^{r \otimes s})^2-(T_A^r)^2 
\otimes {\Bbb I} - {\Bbb I} \otimes (T_A^s)^2].
\end{equation}  
Combining eqs.(\ref{D0},\ref{cas}) we arrive at 
  \begin{equation} \label{D}
D =\frac{1}{2} \sum_{r,s} \sum_{j \in r \otimes s}
(C_j-C_r-C_s)|\psi_j(\phi_r \otimes \phi_s)|^2, 
\label{1} \end{equation}
$\psi_j(\phi_r \otimes \phi_s)$ being the projector of $\phi_r \otimes \phi_s$
onto the irrep $j$ and $C_k$ the Casimir of the irrep $k$. 
The above equation reduces the $D$ term to a sum of square norms of 
irreps  of the gauge group, eqs~(\ref{d1},\ref{d2}) 
hold for each one of the square norms $|\psi_j(\phi_r \otimes \phi_s)|^2$. 
If $\rho$ is free of gravitational anomalies
then 
$0 = \tr (T_A^r \otimes T_A^s) = 
\sum_{j \in r \otimes s} \text{dim}(j) (C_j-C_r-C_s)$. This implies 
that  some 
of the coefficients $(C_j-C_r-C_s)$ in (\ref{D}) are negative. 
In example \ref{gauge}.1
 the only such term corresponds  to a $G^c$ singlet and 
 $D$ is readily seen to be convex along any $\exp(-sT)\phi$ curve.\\

\noindent
{\it Example \ref{gauge}.2:} Consider $G=SO(N)$ with a single vector field,  
$\rho \otimes_s \rho$ contains a symmetric tensor (for which 
$C-2C_{\rho}$ is positive), and a $G^c$ singlet. In this example again, 
the only negative coefficient in eq.~(\ref{D}) 
accompanies a $G^c$ singlet, for any $\phi$ and $T$ $D(\exp(-sT)\phi))$ 
is convex,  
nsv occur only in closed $G^c$ orbits, and Theorem II applies in 
$\widehat{\c^{N}} 
= \c^{N}$.\\

\noindent
{\it Example \ref{gauge}.3:} In $N_F$ flavor, $N$ color SQCD (\ref{D})
contains symmetric and adjoint tensors, for which $C > 2C_{fund}$, 
some $G^c$ singlets and antisymmetric tensors, for which $C < 2C_{fund}$.
In the special case $N_F=1$ there is no antisymmetric tensor, $D(s)$ 
is convex and Theorem II holds. For larger $N_F$ a more detailed 
analysis is required. 
Consider, e.g, 
the case $N_F=2, N=3$ and  the configuration point $\phi_0=(Q^{\a}_i, 
{\tilde{Q}}^j_{\b})$ given by   $Q^{\a}_1=(x,y,0)$, 
$Q_2^{\a}=(u,0,0)$, $\tilde{Q}_{\a}^j=0.$ As $\phi_0 \neq 0$ and 
$\hat{\phi} (\phi_0)=0$,  
$G^c \phi$ is non-closed. Eq.~(\ref{D}) yields $D \propto
(N-1)(|Q_1|^4+|Q_2|^4+|Q_1|^2|Q_2|^2+|Q_1^{\dagger} Q_2|^2)-
(N+1)(|Q_1|^2|Q_2|^2-|Q_1^{\dagger}Q_2|^2)$.  The $SU(3)$ generator $T=\text{diag}(1,1,-2)$ 
is as in Mumford's theorem, and $D(e^{-sT}\phi_0)=D(\phi_0)e^{-2s}$ 
is convex, the exponentially decaying 
terms with negative coefficients in~(\ref{D}) get  cancelled
by positive coefficient terms with the same decaying rate.
For other choices, like $T'=\text{diag}(1,2,-3)$, the negative coefficient 
exponential 
terms persist  but still $D(s)$ is convex. Note that among 
the normalized $\lie G$ generators $\text{diag}(1,1,-2)/\sqrt{6}$ 
is the one that steers $\phi_0$ 
to zero fastest. \\

As this example suggests, to determine the convexity of 
$D(\exp(-sT)\phi)$,  eq.(\ref{D}) should be supplemented with 
information on the weight decomposition $\phi=\sum_{\lambda} \phi_{\lambda}$. 
As $G$ is compact, the $\lambda(T)$'s are rationally related, i.e., 
$\lambda(T) = nq$, $n$ a nonnegative integer, $q$ a ``unit charge". 
The  problem of determining if  $D(s)$ 
is convex reduces 
to a problem of existence of roots of the polynomial $p(x)\equiv D''(s)$,  $x=\exp(-sq)$, 
in the range $0\leq x \leq 1$. The convexity of $D$ along Mumford type 
curves would exclude points in  non-closed $G^c$ orbits from the set of nsv.  
In this case ($G^c \phi$ non-closed and $\exp(-sT)\phi$ as in Mumford's
 theorem) we know that the weight vectors $\lambda$ are all in the half 
space $\lambda(T) \geq 0$, as $\lim_{s \to \infty} \exp(-sT)\phi$ exists. 
For $G$ semisimple, no 
generic result has been  obtained so far regarding the convexity 
of $D(s)$. The analysis is simplified in 
the abelian case $G=U(1)^k$, for which we have a fairly straightforward 
way to determine wether $D(s)$ is convex or not.

\subsection{$U(1)^k$ gauge groups}

From eq.~(\ref{D}), or more directly 
inserting $\phi = \sum_{\lambda} \phi_{\lambda}$ in 
$D(\phi) = \sum_A (\phi^{\dagger}T_A \phi)^2$, $T_A$ an orthonormal basis 
of $\lie G$, we obtain a simple expression for $D$ in the abelian case:
\begin{equation} \label{D2} 
D = \sum_{\lambda \mu A} |\phi_{\lambda}|^2 
|\phi_{\mu}|^2 \lambda(T_A) \mu(T_A) 
= \sum_{\lambda \mu}  <\lambda,\mu> |\phi_{\lambda}|^2 
|\phi_{\mu}|^2, 
 \end{equation} 
from where 
\begin{equation} \label{D3}
D(\exp(-sT \phi)) = \sum_{\lambda \mu}  <\lambda,\mu> |\phi_{\lambda}|^2 
|\phi_{\mu}|^2 e^{-2s(\lambda(T)+\mu(T))}. 
\end{equation}
In the abelian case, we also have a simple criterion to determine 
whether $G^c \phi$ is closed or not: Construct the convex set 
\begin{equation} \label{convex}
S_{\phi} = \left\{ \sum_{{\phi}_{\lambda} \neq 0 } 
 C_{\lambda} \; \lambda \; |\;  0 \leq C_{\lambda} \leq 1 \right\} 
\end{equation} 
It can be shown that: \\
(a) $0$ is outside  $S_{\phi}$ iff $G^c \phi$ is a non-closed 
orbit and $\hp(\phi)=0$, \\
(b) $0$ is a boundary point of 
$S_{\phi}$ iff $G^c \phi$ is a non-closed orbit 
and $\hp(\phi) \neq 0$,\\
(c) $0$ is an inner point of 
$S_{\phi}$ iff $G^c\phi$ is closed. \\
The proof follows trivially from  propositions 5.3 and 6.15 in~\cite{pv}.\\

\noindent
{\it Example \ref{gauge}.4:} 
In a $n-$dimensional  $U(1)$ representation, the weights $\lambda$ 
of a point $\phi_0$ in a non-closed orbit lie all to the right of 
$0$,  all coefficients in~(\ref{D3}) are 
non-negative, $D''(s) > 0$ and, for any superpotential,  the stationary 
points of $V$ lie all on closed orbits. 
This generalizes example \ref{gauge}.1.\\

\noindent
{\it Example \ref{gauge}.5:} Consider the $U(1) \times U(1)$ 4-dimensional 
representation  
with orthonormal  generators 
\begin{equation}
T_1=\frac{1}{2}\left( \begin{array}{cccc} -1&0&0&0\\0&1&0&0\\0&0&1&0\\0&0&0&-1 
\end{array} \right) \;\;\; 
T_2= \frac{1}{2} \left( \begin{array}{cccc}
 1&0&0&0\\0&1&0&0\\0&0&-1&0\\0&0&0&-1 
\end{array} \right)
\end{equation}
The weight diagram is a square centered 
on $0$ (figure 1). The weights are orthogonal, 
the matrix $<\lambda, \mu>$ in eqs.~(\ref{D2}, \ref{D3}) 
is diagonal, $D(\exp(-sT)\phi)$ is 
convex for any $\phi$ and $T$, 
and Theorem II holds in the entire configuration 
space.  Vectors can be  classified according to 
 the number of nonzero weights. There are two classes of vectors 
in  closed orbits:  
(i) 4 weight vectors and  (ii) two opposite weight vectors. There 
are three different types of vectors in non-closed orbits: (iii) three 
weight vectors, which satisfy $\hp(\phi) \neq 0$, 
and (iv) two adjacent weight vectors and (v) one weight vectors, 
for which $\hp(\phi)=0$, i.e., they are in the same fiber as 
$\phi=0$.  Take, e.g, case (iii), Mumford's curve $\phi(s)$
``shuts down" one weight leaving a case (ii) vector. 
The basic invariants are ${\hp}^1=\phi^1 \phi^3$ 
and ${\hp}^2=\phi^2 \phi^4$, they are unconstrained, 
then ${\cal D} = \c^2$. For any $W$,  
${\cal M}_v = {\cal M}_{sv} \cup {\cal M}_{nsv}$ 
will be  a subset of ${\cal D} = \c^2$. \\ 

\noindent 
{\it Example \ref{gauge}.6:} Consider the $U(1) \times U(1)$
 6-dimensional representation 
\begin{equation}
T_1=\frac{\sqrt{3}}{6}\left( \begin{array}{cccccc} 
-1&0&0&0&0&0\\0&1&0&0&0&0\\0&0&2&0&0&0\\0&0&0&1&0&0\\
0&0&0&0&-1&0 \\0&0&0&0&0&-2 
\end{array} \right) \;\;\; 
T_2= \frac{1}{2} \left( \begin{array}{cccccc}
 1&0&0&0&0&0\\0&1&0&0&0&0\\0&0&-1&0&0&0\\0&0&0&0&0&0 \\0&0&0&0&-1&0\\
0&0&0&0&0&0  
\end{array} \right)
\end{equation}
The weight diagram is a hexagon centered on $0$ (figure 2). 
By excluding  two adjacent weights we get a 4-weight 
vector 
in a non-closed orbit. Take e.g $\phi_0=(\phi^1,\phi^2,\phi^3,\phi^4,0,0), 
\phi^i \neq 0, i = 1,...,4$, the boundary of $S_{\phi}$ 
appears in dotted lines in figure 2. It can readily be 
checked that: (1) there is a unique choice of $T$ 
satisfying   Mumford's theorem, (2) $e^{-sT}\phi_0$ turns off 
$\phi^2$ and $\phi^3$ and (3)  $d^2D/ds^2$ may (i) 
change sign, (ii) be positive, (iii) be negative, and that $D(s)$ 
may even grow along this curve depending on the values 
of the  $\phi^i$'s. Theorem II does {\em not} 
apply, we cannot 
draw any conclusions for this theory.  \\

\subsection{Energy bounds in core-to-core theories}

There are many  examples of theories for which $\partial W(\cdot)$ 
sends the core of 
\df points in  closed $G^c$ orbits in $\c^n$ 
onto the core of *\df points of  closed orbits 
in ${\c^n}^*$. For these theories, given any 
point $\phi_0$ in a non-closed $G^c$ orbit, the \df points in 
the closed orbit in the boundary of 
$G^c \phi_0$ have lower energy.\\

\noindent
{\bf Theorem III:} 
Assume  $\partial W(\cdot)$ sends \df points onto *\df points,  
i.e. $[\partial W(\phi)] T [\partial W(\phi)]^{\dagger}=0 \; \forall 
T \in \lie G$  
if  ${\phi}^{\dagger}  T \phi = 0 \; \forall T \in \lie G$. 
If $G^c \phi_0$ is non-closed and $\phi_D$ is a 
\df point in the boundary of $G^c \phi_0$, then $V(\phi_D) < V(\phi_0)$.\\
Proof: Let $\phi_c$ be as in Mumford's theorem, 
$\phi_D$ a \df point in the closed orbit $G^c \phi_c$. 
As $\partial W (\phi_D)$ is *\df, $\phi_D$ is a global minimum 
of the restriction of $F$ to $G^c \phi_c$, then $F(\phi_D) \leq F(\phi_c)$. 
As $F$ decreases along Mumford's curve $F(\phi_c) \leq F(\phi_0)$.
Thus $F(\phi_D) \leq F(\phi_c) 
\leq F(\phi_0)$, and also $0 = D(\phi_D) < D(\phi_0)$, from where  
 $V(\phi_D) < V(\phi_0)$. \\

\noindent
{\it Example \ref{gauge}.7:} Theories having a single 
basic invariant satisfy the hypothesis of Theorem~III 
(see example~\ref{global}.4). Table \ref{tab:1} lists 
all asymptotically free, anomaly free representations of simple groups 
having a single basic invariant, they were obtained from \cite{dms}. 
For all these theories $V(\phi_D)$, $G\phi_D$  
the core of \df points in the boundary of 
the  non-closed orbit $G^c \phi_0$, gives a lower bound 
to the energies $\{ V(\phi) | \phi \in G^c \phi_0 \}$ 
Among these representations, the real ones have the property 
that, for any invariant $W$,  $\partial W(\phi) (-T) 
(\partial W (\phi))^{\dagger} \propto \phi^{\dagger}T \phi$
 (example~\ref{global}.4), this implies that \df points 
satisfy the $MD-$flat condition eq.(\ref{m}).  
For a subset of the real $\rho$'s in Table~\ref{tab:1} 
the tensor decomposition 
$\rho \otimes_s  \rho$ contains only two irreps, one of which  is a singlet, 
For them, theorem~II holds in the entire configuration space, 
and, as happens for SQCD, the stationary points of $V$ are $D-$flat, 
a non generic feature among the theories satisfying the hypothesis of 
Theorem~II.

\begin{table}
\caption{\label{tab:1} 
All anomaly free representations of simple groups $G$ with 
a single basic holomorphic $G$ invariant. Entries 1-14 satisfy the 
hypothesis of Theorem~III, entries 1,3,5,6 and 12 also 
satisfy the hypothesis of Theorem~II. Pseudo-real representations are 
{\em not} checked in the fourth column, real representations are 
required in order that $(\partial W) T (\partial W)^{\dagger} 
\propto \phi^{\dagger}T \phi$. 
In the last column Dynkin labels are used to avoid complicated Young diagrams.}
\setlength{\tabcolsep}{15mm}
\renewcommand{\arraystretch}{1.5}
\[
\begin{array}{|c|c|c|c|c|} \hline
& G & \rho &  \text{real} & \rho \otimes_s \rho \\ \hline \hline
1& SU(N) & \Yfund + \overline{\Yfund} & \surd & 
\Ysymm + \overline{\Ysymm} + Adj + {\Bbb I}    \\ 
2 & SU(6) & \Ythreea & & [0,0,2,0,0] + adj     \\
3 & SU(4)  &  \Yasymm & \surd & [0,2,0] + {\Bbb I} \\
4 & SU(2) & \Ythrees  & & [2] + [6]     \\
5 & SO(N)  & \Yfund & \surd & \Ysymm + {\Bbb I}  \\
6 & SO(7),  & spinor & \surd & [0,0,2] + {\Bbb I}    \\
7 & SO(9)   & spinor & \surd & \Yfund + [0,0,0,2] + {\Bbb I}  \\
8 & SO(N), N=11,12,14 & spinor  & &  [0,...,0,2] + \left[
\Yasymm + \Yfund , \; 
\Yasymm , \;\Ythreea \right] \\
9 & SO(10)  & 2 \; spinors &  & 3[0,0,0,0,2]+[0,0,1,0,0]+{\Bbb I} \\
10 & Sp(2N)   & \Yfund +  \Yfund &  \surd  & 3[2,0,0,...,0] + [0,1,0,...,0] + 
{\Bbb I}   \\
11 & Sp(6)  & \Ythreea  &  &  [2,0,0]+[0,0,2]  \\
12 & G_2  &  {\bf 7}   & \surd & [2,0]+ {\Bbb I}   \\
13 & E_6   &  {\bf 27}  &  &  [2,0,0,0,0,0]+[1,0,0,0,0,0]  \\
14 & E_7  &  {\bf 56}  &     &  [2,0,0,0,0,0,0]+[1,0,0,0,0,0,0]  \\ 
\hline
\end{array} \]
\end{table} 

There are many other examples of theories for which $\partial W(\cdot)$ 
sends \df points onto *\df points. Theorem III applies for all these theories.\\

\noindent 
{\it Example \ref{gauge}.8 :}
 For $N_F<N$ ($N_F=N$) the basic SQCD  holomorphic invariants 
are $M_i^j = Q_j^{\a} \q_{\a}^i$ (and $B = \det Q, \tilde{B} = 
\det \tilde{Q}$). A 
straightforward calculation shows that 
the gradient of any flavor invariant superpotential $W(\text{det} M)$ 
sends \df points onto *\df points.\\

\section{Conclusions} 
We proved in Theorems~I and II that  
for a large set of theories with a 
compact global symmetry $G$ and gauge theories with gauge group $G$, every non-supersymmetric vacuum is \df or $G^c$ related to a \df point. This not 
only simplifies the search of nsv but also leads to a parametrization of its 
moduli space ${\cal M}_{nsv}$ in terms of basic holomorphic invariants, 
extending the well known technique 
of constructing ${\cal M}_{sv}$. 
We also showed in Theorem~I that in generic theories with a compact global  
symmetry $G$, if $G^c \phi_0$ is non-closed, 
 a lower energy point exists in the closed $G^c$ orbit in the boundary of 
$G^c \phi_0$. This is also the case 
for a number of gauge theories, for which a \df point in the boundary 
of a non-closed orbit $G^c \phi_0$ always has lower energy than $\phi_0$ 
(Theorem III). To our knowledge, 
these  are the  first known results  
on moduli spaces of non-supersymmetric vacua. They  uncover  
an unexpected  connection between non-supersymmetric vacua 
and the $D-$flatness condition.\\

\section{Acknowledgements}
I would like to thank M.~L.~Barberis, L.~Cagliero and W.~Skiba for 
helpful discussions, W.~Skiba for useful comments on the 
manuscript  that led to an improved organization. This work was supported by Fundaci\'on Antorchas.

\vspace{1cm}

\setlength{\unitlength}{1mm}
\begin{picture}(40,40)(0,0)
\put(0,20.85){\vector(1,0){44}}
\put(20.85,0){\vector(0,1){40}}
\put(40,24){$\lambda(T_1)$}
\put(22,37){$\lambda(T_2)$}
\put(30,30){${\bullet}_{\lambda_2}$}
\put(30,10){${\bullet}_{\lambda_3}$}
\put(10,30){${\bullet}_{\lambda_1}$}
\put(10,10){${\bullet}_{\lambda_4}$}
\put(46,48){\makebox(0,0){Figure 1: weight diagram for the theory of 
Example~\ref{gauge}.5.}}
\end{picture}
\vspace{2cm}

\begin{picture}(40,40)(0,0)
\put(46,48){\makebox(0,0){Figure 2: weight diagram for the 
theory of Example~\ref{gauge}.6}}
\put(0,20.85){\vector(1,0){44}}
\put(20.85,0){\vector(0,1){40}}
\put(40,24){$\lambda(T_1)$}
\put(22,39){$\lambda(T_2)$}
\put(28.5,35){${\bullet}_{\lambda_2}$}
\put(37,20){${\bullet}_{\lambda_3}$}
\put(28.5,5){${\bullet}_{\lambda_4}$}
\put(11.5,5){${\bullet}_{\lambda_5}$}
\put(3,20){${\bullet}_{\lambda_6}$}
\put(11.5,35){${\bullet}_{\lambda_1}$}
\dottedline(12.3,36)(29.3,6)
\dottedline(12.5,36)(29.5,36)
\dottedline(29.5,36)(38,21)
\dottedline(38,21)(29.5,6)
\end{picture}

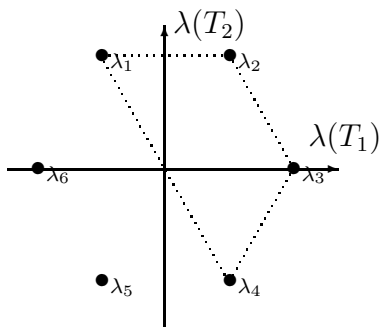
\end{document}